# A comparative theoretical study on physical properties of synthesized $AVO_3$ (A = Ba, Sr, Ca, Pb) perovskites

Khandaker Monower Hossain[1], Mirza H. K. Rubel[1,*] M. M. Rahaman[1,2,*] M. M. Hossain[3], Md Imran Hossain[1], Anjuman Ara Khatun[1], J. Hossain[4] and A. K. M. A. Islam[5,6]

[1]Department of Materials Science and Engineering, University of Rajshahi, Rajshahi–6205, Bangladesh

[2]Geophysical Laboratory, Carnegie Institution for Science, Washington, DC 20015, USA

[3]Department of Physics, Chittagong University of Engineering and Technology, Chittagong–4349, Bangladesh

[4]Department of Applied Physics and Electronic Engineering, University of Rajshahi, Rajshahi–6205, Bangladesh

[5]Department of Physics, Rajshahi University, Rajshahi–6205, Bangladesh

[6]International Islamic University Chittagong, 154/A College Road, Chittagong–4203, Bangladesh

*__Corresponding author:__ Mirza H. K. Rubel; Md. Mijanur Rahaman

Cell: +8801714657365; E–mail: mhk_mse@ru.ac.bd or mirzamse@gmail.com; mijan_mse@ru.ac.bd

**Abstract:** In this paper, we employ CASTEP based on DFT (density functional theory) calculations to investigate various physical properties of $BaVO_3$, $SrVO_3$, $CaVO_3$ and $PbVO_3$. The elastic constants, bulk modulus, Shear modulus, Young's modulus, Pugh's ratio, Poisson's ratio, Vickers hardness, universal anisotropy index and Peierls stress are calculated to rationalize the mechanical behavior of the aforementioned compounds. The study of electronic band structure and density of states (DOS) reveal the strong evidence of metallic behavior for all the perovskites. The analysis of bonding properties exhibits the existence of covalent, ionic and metallic bonds. The optical properties of $AVO_3$ have been carried out and are discussed in this paper as well. The analysis of phonon property implies the dynamical stability of $BaVO_3$ but not for $SrVO_3$, $CaVO_3$ and $PbVO_3$. The values of Debye temperature and minimum thermal conductivity imply that only $PbVO_3$ compound has potential to be used as TBC material.

**Key–words:** AVO$_3$ perovskites, DFT calculations, mechanical properties, electronic band structure, optical properties, phonon spectra and thermodynamic properties.

---

## 1. Introduction

The ABO$_3$–type perovskite oxides and its derivatives bear significant importance, because of their structures and properties can be modified by the substitution of different elements into their crystal sites. These interesting classes of materials are technically important and carry fundamental interest in the field of materials science, physics, chemistry, and microelectronics [1] as well. In the midst of them, the AVO$_3$ (A = Sr, Ba, and Pb) type perovskite–based oxides reveal interesting properties and have many potential applications as photovoltaics, displays and solid-state lighting, smart windows, multiferroic devices, ferroelectric devices, high–temperature solid oxide fuel cells, high–*T*c superconductor, and so on [2,3]. Among these multiferroic property in which ferromagnetism and ferroelectricity coexist, have gathered huge experimental and theoretical interest owing to their considerable technological and fundamental importance [4–8]. As a part of experimental endeavor, four oxide compounds such as BaVO$_3$ [9], SrVO$_3$ [10], CaVO$_3$ [11] and PbVO$_3$ [12] have been synthesized by using high temperature and high pressure (HTHP) conditions to investigate their structural and relevant properties. The primal phase of PbVO$_3$ was tetragonal perovskite (*P4mm*) (*T*–phase) was obtained at high pressure is a candidate multiferroic with a two–dimensional *C*–type antiferromagnetism (*C*–AFM) ordering and a large ferroelectric polarization [13–15]. Moreover, the *T*–phase of PbVO$_3$ is isostructural with PbTiO$_3$ ferroelectric but exhibits a more pronounced structural distortion with a larger unit cell volume. However, ABO$_3$–type ferroelectric (FE) perovskite compounds often transform to a paraelectric (PE) phase with the change of temperature and/or pressure [16–18]. For instances, the AVO$_3$ and ATiO$_3$ tetragonal perovskites (*P4mm*) transform to cubic perovskite ($Pm\bar{3}m$) at high temperature and pressure [12,14,19–21]. Furthermore, first–principles calculations on PbVO$_3$ perovskites [22] were carried out to observe conductivity and the pressure–induced phase transition in order to elucidate and complement the experimental data. The first–principle calculations on cubic phase of *C*–PbVO$_3$ predicted the compound as nonmagnetic metal along with metallic conductivity. Another compound BaVO$_3$ showed pure polycrystalline phase with cubic perovskite structure under high-pressure treatment and it should be noted that if either the pressure or temperature was insufficient Ba$_3$(VO$_4$)$_2$ was formed instead of BaVO$_3$ [9]. Notably, SrVO$_3$ and CaVO$_3$ are examples of two correlated Mott conductors with perovskite structure and their oxygen nonstoichiometry effects on the structures and electronic states are investigated extensively [2,10,11]. Between them, Strontium vanadate [23] has a cubic perovskite structure with $a$ = 3.842 Å [2] that exhibits a Pauli paramagnetic and metallic character near room temperature as well. Additionally, the dielectric function, band structure, Fermi surface, interband optical transitions of $SrVO_3$ transparent conducting thin film has also been analyzed from first–principles study [2]. Moreover, this material has achieved great attention as a potential oxide electrical conductor [24,25]. Recently, double–perovskite Bi oxides are a new series of magnetic materials showing superconductivity; in the midst of them the novel A–site ordered double perovskite [26–28] and simple perovskite [29,30] bismuthates have been reported. Their first principles calculations are also reported [31,32] to predict various interesting physical properties such as electronic, mechanical and thermodynamic properties. In addition, most recently, the lattice dynamic and their related physical properties of $ABO_3$–type materials have been investigated by inelastic light scattering [33–35]. Furthermore, there are many examples of single–phase materials such as $BiFeO_3$, $PbVO_3$, $BiMnO_3$, $CrBiO_3$, $BiCoO_3$ etc. exhibiting multiferroic behavior [7,36,37].

Herein, the cubic perovskite structures of $AVO_3$ is our point of interest showing the analogy with the previously reported ferroelectric oxides $BaTiO_3$, $CaTiO_3$, $PbTiO_3$, $SrTiO_3$ [38–40]. The A (= Ba, Sr, Pb, Ca) and V atoms were located on the site 1$a$ (0, 0, 0) and 1$b$ (0.5, 0.5, 0.5) respectively, following the $ABO_3$ cubic perovskite crystallographic structure. The oxygen atoms were set on the site 3$c$ (0.5, 0, 0.5). The X–ray powder diffraction pattern of $AVO_3$ were indexed with the cubic cell with the lattice parameters ($a$ = 3.9428 Å, $a$ = 3.8425 Å, $a$ = 3.7860 Å and $a$ = 3.8361 Å for $BaVO_3$, $SrVO_3$, $CaVO_3$ and $PbVO_3$ respectively). Moreover, the magnetic properties of these compounds have been discussed earlier. Nowadays, the cubic phases of $AVO_3$ and $ATiO_3$ materials are an extensive research area for the proper description of electronic, elastic, thermodynamic, vibrational and optical properties for the promising applications. Recently, M. Kamruzzamanan *et al.* [41] reported a comparative study on $ATiO_3$ based on first principle calculations. In order to quantify the ferroelectric instability of the cubic phase of $ATiO_3$, the phonon frequencies and Born effective charges have also been determined by calculating the interatomic forces for several small displacements consistent with the symmetry of modes. Moreover, the specific heat of $ATiO_3$ was calculated as a first report. But theoretical information is not available yet as on elastic, electronic, optical, population analysis, phonon spectra and thermodynamic properties of $AVO_3$ compounds. Notably, first principle calculations are well–known method to predict the interesting phenomena such as mechanical, electronic, optical, thermodynamic properties and so on to understand a new material system precisely.

Therefore, our attempt is to calculate and investigate several microscopic and macroscopic attributes of $AVO_3$ materials based on DFT calculations. In this research paper, we calculate and compare the structural, mechanical (elastic constants, bulk, shear and Young's modulus, Pugh's and Poisson's ratio, Peierls stress, Vickers hardness and Cauchy pressure), electronic (band structure, DOS, charge density map, Fermi surface), optical (dielectric function, dielectric loss, refractive index, photoconductivity, absorbance and reflectance), population, phonon and thermodynamic properties (Helmholtz free energy, energy, entropy, specific heat, Debye temperature, melting temperature and minimum thermal conductivity) of $AVO_3$ materials employing CASTEP–Code software package.

## 2. Computational method

The Cambridge Serial Total Energy Package (CASTEP) code [42–45] for the density functional theory (DFT) calculations [46,47] is applied in the present work. The norm–conserving pseudopotential scheme is also utilized to model the interactions of valence electrons with ion cores. The electronic exchange correlation interaction is treated under the local density approximation (LDA) within the scheme described by Ceperley–Alder (CA) [48] and Perdew–Zunger (PZ) functional [49] which depends on both the electron density and its gradient at each space point. The well–known Broyden–Fletcher–Goldfarb–Shanno (BFGS) variable–metric minimization algorithm [50] is then utilized to seek the ground state or geometry optimization. It has been demonstrated that it is a very efficient and robust means to explore the optimal minimal energy crystalline structure. The plane wave basis set is truncated using a cutoff energy of 700 eV. To sample the Brillouin zone [51], the Monkhorst–Pack $k$–point mesh is employed [52]. It is selected based on the convergence of the $k$–point mesh, where the change of total energy becomes less than 1 meV/atom [53–56]. In the study, the selected $k$–point mesh is 11x11x11. The convergence conditions considered in the geometry optimization include an energy tolerance of $5 \times 10^{-6}$ eV/atom, maximum ionic Hellmann–Feynman force within 0.01 eV/Å, maximum stress within 0.02 GPa and maximum ionic displacement within $5 \times 10^{-4}$ Å. These parameters were very convenient to lead good convergence of total energy and internal forces.

## 3. Results and discussion

### 3.1. Structural Properties

At first the crystal structure of AVO$_3$ materials were drawn using the available refinement data [9–12]. The oxides AVO$_3$ crystallize in a cubic perovskite structure with space group of $Pm\bar{3}m$ (No. 221) consisting of A (= Ba, Sr, Ca and Pb) atom at the corner, V in the body centre and O at the face centre of the cube. The structures are fully relaxed by optimizing the geometry with the lattice parameters and internal coordinates. The crystal structures of BaVO$_3$, SrVO$_3$, CaVO$_3$ and PbVO$_3$ as a structural model of cubic AVO$_3$ system are depicted in Fig. 1. The optimized lattice parameters of AVO$_3$ compounds both experimental and theoretical are summarized in Table 1. The calculated values match reasonably with the experimental result [9–12].

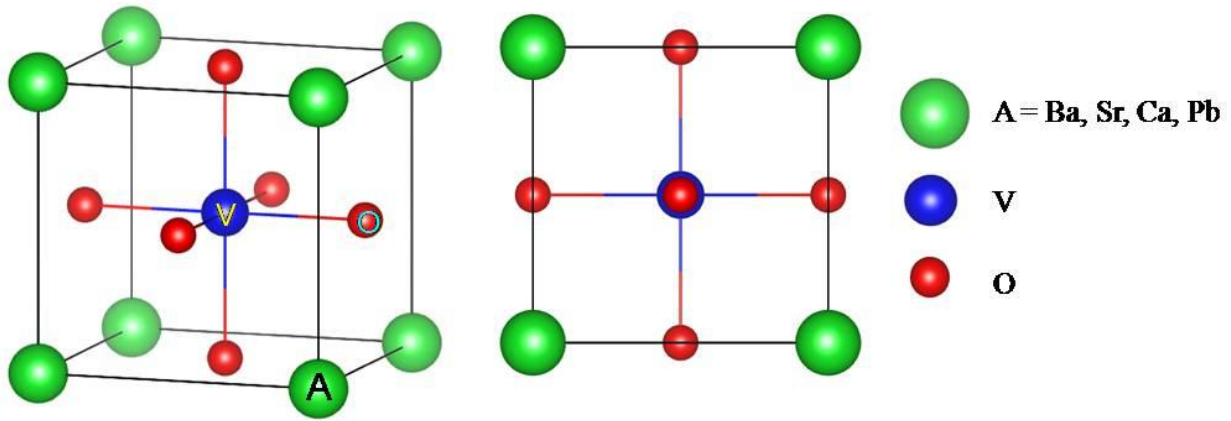

Fig. 1. (Color online) Crystal structure of perovskite AVO$_3$ (A = Ba, Sr, Ca, Pb), (a) Unit cell, and (b) 2D view in xz plane.

Table 1. Calculated lattice constants ($a$) in Å and Volume ($V$ in Å$^3$) of AVO$_3$ compounds.

| Compounds | Lattice constant, $a$ (Experimental) | Lattice constant, $a$ (Present study) | Volume, $V$ (Present study) |
|---|---|---|---|
| BaVO$_3$ | 3.9428(3) [9] | 3.9090 | 59.7333 |
| SrVO$_3$ | 3.8425(1) [10] | 3.8269 | 56.0486 |
| CaVO$_3$ | 3.7860(3) [11] | 3.7803 | 54.0230 |
| PbVO$_3$ | 3.8361(4) [12] | 3.8867 | 58.7163 |

## 3.2. Mechanical Properties

The study of mechanical properties such as elastic moduli, ductile/brittle behavior, and elastic anisotropy are of critical importance for industrial applications of engineering materials. The elastic constants are obtained from linear finite strain–stress method within the CASTEP–code [57]. An elastic property is the measurement of the tendency of a solid to deform non–permanently in various directions with applied stress.

The Born stability criteria, which define the mechanical stability of a lattice, are typically formulated in terms of the $C_{ij}$ [58,59]. In this case, they are valid only in the limit of vanishing of pre−stress. Several studies [60,61] have demonstrated that the appropriate stability criteria for a stress lattice are those which are formulated in terms of the stress−strain coefficients ($C_{ij}$) and hence are based on enthalpy considerations. The three stability criteria for a cubic crystal are $C_{11} + 2C_{12} > 0$, $C_{44} > 0$, $C_{11} - C_{12} > 0$ which are referred to as spinodal, shear, and Born criteria, respectively. The spinodal criterion is equivalent to requiring that the bulk modulus be positive. Thus, the mechanical stability in a cubic crystal leads to the restrictions on the elastic constants that B, $C_{11} - C_{12}$ and $C_{44}$ should be positive [62]. The elastic constants presented in Table 2 obey these stability conditions showing that $AVO_3$ is mechanically stable. The $CaVO_3$ perovskite has the largest single crystal elastic constants ($C_{11}$) compared to others. It is also noticed that the calculated value of $C_{11}$ for all perovskites is higher than $C_{12}$, suggesting that the incompressibility along crystallographic a−axis is stronger than that along b−axis. This means that the bonding strength along the [100] direction is higher than that along the [011] directions in these compounds.

The bulk modulus (*B*) and shear modulus (*G*) are also calculated from the single crystal zero−pressure elastic constants through the Voigt−Reuss−Hill formula [63,64]. In addition, the Poisson's ratio (*v*) and the Young's modulus (*Y*) can also be obtained using well−known relationships [65]. All these calculated elastic parameters are summarized in Table 2. The bulk modulus evaluates the average bond strength of constituent atoms for a given solid [66]. Recently, M. T. Nasir *et al.* [67] reported the strong bonding in ScIrP and ScRhP by the bulk modulus (190 and 171 GPa). Therefore, the calculated values of *B* (196, 187, 200 and 183 GPa for $BaVO_3$, $SrVO_3$, $CaVO_3$ and $PbVO_3$ respectively) may reflect the strong bonding strength of atoms involved in these compounds. The bond strength of atoms also gives the required resistance to volume deformation under external pressure. On the other hand, change of shape in a solid crucially depends on its shear modulus *G*, which shows a crucial correlation with material's hardness. The bulk and shear modulus do not measure the material hardness directly, Vickers hardness can be estimated by Chen's formula [68], given by: $Hv = 2(k^2G)^{0.585} - 3$, where, $k(=G/B)$ is the Pugh's ratio. The calculated values of the Pugh's ratio (*k*), and Vickers hardness (*Hv*), Poisson's ratio (*v*), and Cauchy pressure are presented in Table 2. The Young's modulus (*Y*) exerts influence on the thermal shock resistance of a solid. The critical thermal shock coefficient varies inversely to the Young's modulus (*Y*) [69]. The greater the value of thermal shock coefficient, the better the thermal shock resistance. A material is selected as a thermal barrier coating (TBC) substance depending on thermal shock resistance. Notably, PbVO$_3$ has comparatively lower Young's modulus than the materials under this investigation and hence it might be suitable as TBC material.

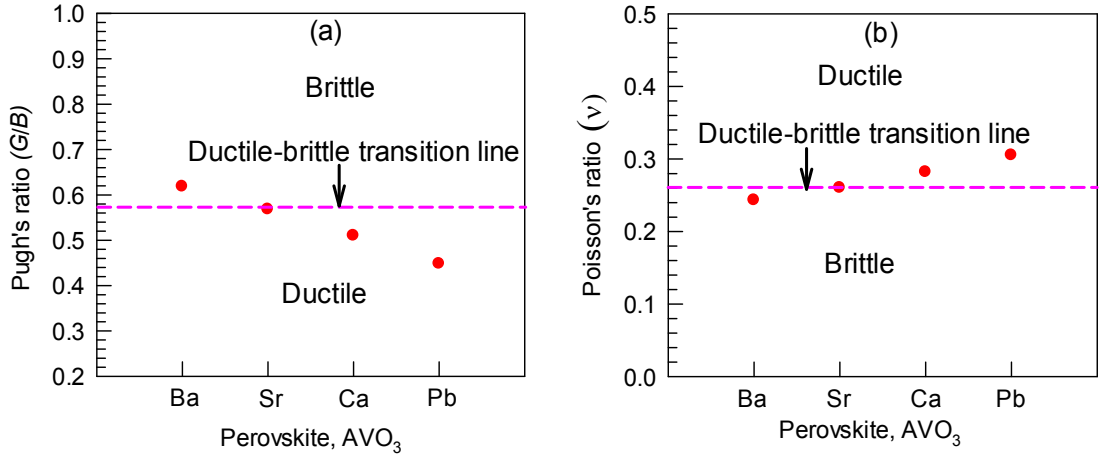

Fig. 2. (a)The calculated Pugh's ratio ($G/B$) and (b) Poisson's ratio ($v$) show the graphical representation of ductile/brittle behavior of the considered perovskite, AVO$_3$ (A = Ba, Sr, Ca and Pb). The horizontal dashed line indicates the ductile-brittle transition line.

The study of failure mode, *i.e.*, brittle or ductile nature of a material is technologically important. For most practical situations, a material may be classified as either ductile or brittle. Fig. 2 shows the graphical representation of ductile materials from brittle materials for all considered perovskite materials. A material is said to be ductile if the value of $k$ is smaller than 0.57 [70]; otherwise, the material is brittle. As can be seen in Table 2 and Fig. 2(a), the BaVO$_3$ is surely brittle material whereas the brittle/ductile nature of SrVO$_3$ is at ductile−brittle transition line; on the other hand, CaVO$_3$ and PbVO$_3$ exhibit ductile nature. In addition the Pugh ratio, Frantsevich's also proposed a critical value of Poisson's ratio ($v \sim 0.26$) for separating the ductile and brittle nature of solids [71]. The calculated values of Poisson's ratio ($v$) are 0.243, 0.260, 0.282 and 0.305 respectively for BaVO$_3$, SrVO$_3$, CaVO$_3$ and PbVO$_3$ [Table 2 and Fig. 2(b)], again confirming that all considered compounds show almost similar behavior in nature. Moreover, another indicator of failure mode of materials is the Cauchy pressure, defined as ($C_{12}−C_{44}$) [72]. If it is negative, the material is expected to be brittle; otherwise (having positive Cauchy pressure), it is a ductile one [72]. Therefore, the compounds BaVO$_3$ and SrVO$_3$ in this study are assumed to be brittle, besides, CaVO$_3$ and PbVO$_3$ are ductile in accordance with the above mentioned three indicators.

The universal anisotropic index $A^U$, another important quantity, is defined as $A^U = 5\frac{G_V}{G_R} + \frac{B_V}{B_R} - 6 \geq 0$, where $A^U = 0$ is for isotropic materials, and the deviation from zero defines the elastic anisotropy of crystals [73]. The values of $A^U$ for BaVO$_3$, SrVO$_3$, CaVO$_3$ and PbVO$_3$ are 0.042, 0.010, 0.134 and 0.005 respectively implying that all the compounds are anisotropic but PbVO$_3$ indicates reasonably negligible anisotropy.

In this study, the parameter Peierls stress denoted as $\sigma_p$ is the force required to move a dislocation inside an atomic plane in the unit cell. This is an important quantity that measures the strength of a crystal through displacing dislocations. The progress of a dislocation in a glide plane of the synthesized cubic perovskite crystals [9−12] can be estimated using Peierls stress ($\sigma_p$) [74]. By using the shear modulus $G$ and Poisson ratio $v$, Peierls stress can be computed as follows:

$$\sigma_p = \frac{G}{1-v} \exp\left[-\frac{2\pi d}{b(1-v)}\right]$$

Herein, $b$ is the Burgers vector and $d$ is the interlayer distance between the glide planes. The calculated Burger's vector $b$, interlayer distance $d$, and the resulting Peierls stress $\sigma_p$ of AVO$_3$ compounds are also listed in Table 2. The highest and the lowest values of estimated Peierls stress are 2.394 and 1.284 respectively, for BaVO$_3$ and PbVO$_3$. These values of $\sigma_p$ are comparable to the reported double perovskites (Na$_{0.25}$K$_{0.45}$)Ba$_3$Bi$_4$O$_{12}$ [31] and (K$_{1.00}$)(Ba$_{1.00}$)$_3$(Bi$_{0.89}$Na$_{0.11}$)$_4$O$_{12}$ [32], respectively. The $\sigma_p$ of AVO$_3$ compounds may also be compared with several inverse perovskites Sc$_3$InX (X = B, C, N) [75] and some MAX phases Ti$_2$AlC, Cr$_2$AlC, Ta$_2$AlC, V$_2$AlC, and Nb$_2$AlC which attain $\sigma_p$ in the ranges 0.7–0.98 (GPa) [75,76]. Furthermore, the reported $\sigma_p$ for rocksalt binary carbides TiC, VC and CrC have values between 17.46 and 22.87 GPa [76] and exhibiting the sequence, $\sigma_p$ (selected inverse perovskites and MAX phases) < $\sigma_p$ (AVO$_3$ perovskites) << $\sigma_p$ (binary carbides). Therefore, it is explicit that dislocations can move easily in the selected MAX phases, but this is quite impossible in the case of the binary carbides. Since, the AVO$_3$ perovskites in the present study exhibit an intermediate values of $\sigma_p$ larger than MAX phases, dislocation movement may still occur here, but not as easily as in MAX phases. As the value of $\sigma_p$ for PbVO$_3$ is smaller than BaVO$_3$, SrVO$_3$ and CaVO$_3$, dislocation movement in PbVO$_3$ may occur more easily than other perovskites under this study.

Table 2. The elastic constants, $C_{ij}$ (GPa), bulk modulus, $B$ (GPa), shear modulus, $G$ (GPa), Young's modulus, $Y$ (GPa), Pugh's ratio, $G/B$, Poisson's ratio, $v$, Vickers hardness, $H_v$ (GPa), Cauchy pressure, universal anisotropic index $A^U$, Burger's vector $b$ (Å), interlayer distance $d$ (Å) and Peierls stress $\sigma_P$ of $AVO_3$ compounds.

| Compounds | $C_{11}$ | $C_{12}$ | $C_{44}$ | $B$ | $G$ | $Y$ | $G/B$ | $v$ | Cauchy pressure | $H_v$ | $A^U$ | $b$ | $d$ | $\sigma_P$ |
|---|---|---|---|---|---|---|---|---|---|---|---|---|---|---|
| $BaVO_3$ | 323 | 118 | 124 | 186 | 115 | 286 | 0.61 | 0.243 | -6 | 15.28 | 0.042 | 3.909 | 1.954 | 2.394 |
| $SrVO_3$ | 356 | 108 | 118 | 197 | 112 | 284 | 0.56 | 0.260 | -10 | 13.30 | 0.010 | 3.826 | 1.913 | 2.168 |
| $CaVO_3$ | 367 | 117 | 89 | 200 | 102 | 262 | 0.51 | 0.282 | 28 | 10.61 | 0.134 | 3.780 | 1.890 | 1.787 |
| $PbVO_3$ | 289 | 130 | 84 | 183 | 82 | 214 | 0.44 | 0.305 | 46 | 7.29 | 0.005 | 3.886 | 1.943 | 1.284 |

*3.3. Electronic Properties*

*3.3.1. Band structure and Density of states*

The electronic properties of a compound such as the band structure, total density of states (TDOS) and partial density of states (PDOS) are very convenient to elucidate bonding nature and other relevant properties of materials. The result of band structure calculations along the highly symmetric directions within the $K$−space for $AVO_3$ are presented in Fig. 3, where the horizontal dashed line is shown as the Fermi level, $E_F$. The purely valence and conduction bands are indicated with black line, whereas, the bands crossing the $E_F$ with various colored lines. We observe that from the band diagram, several dispersive bands cross the $E_F$ especially along the X−R direction and the valance and conduction band appreciably overlap with each other resulting no band gap at the $E_F$ indicates the products in this report exhibit metallic conductivity. Interestingly, the band structures of $AVO_3$ in the present work are nearly analogous to those reported in the previous articles [31,32,67,77,78].

In order to well understanding the contributions of different orbital to the band formation of $AVO_3$ compounds, total density of state (DOS) and partial density of states (PDOS) as well as the atomic contributions at the $E_F$ have been calculated and are plotted in Fig. 4. It is seen that the TDOS around the $E_F$ arises mostly from V−3$d$ and O−2$p$ for all phases. Therefore, electronic charge should transfer from 3$d$ to 2$p$ owing to the interactions between them. So, there is strong hybridization between V−3$d$ and O−2$p$ for all these compounds. Thus, a strong binary O−V bond in $AVO_3$ results due to this strong hybridizations. Notably, the total DOS at $E_F$ in this report is 1.42, 1.29, 1.24 and 1.77 states/eV/unit cells for $BaVO_3$, $SrVO_3$, $CaVO_3$ and $PbVO_3$ respectively. The value of non−zero DOS at $E_F$ is also an indication of conductive nature of them.

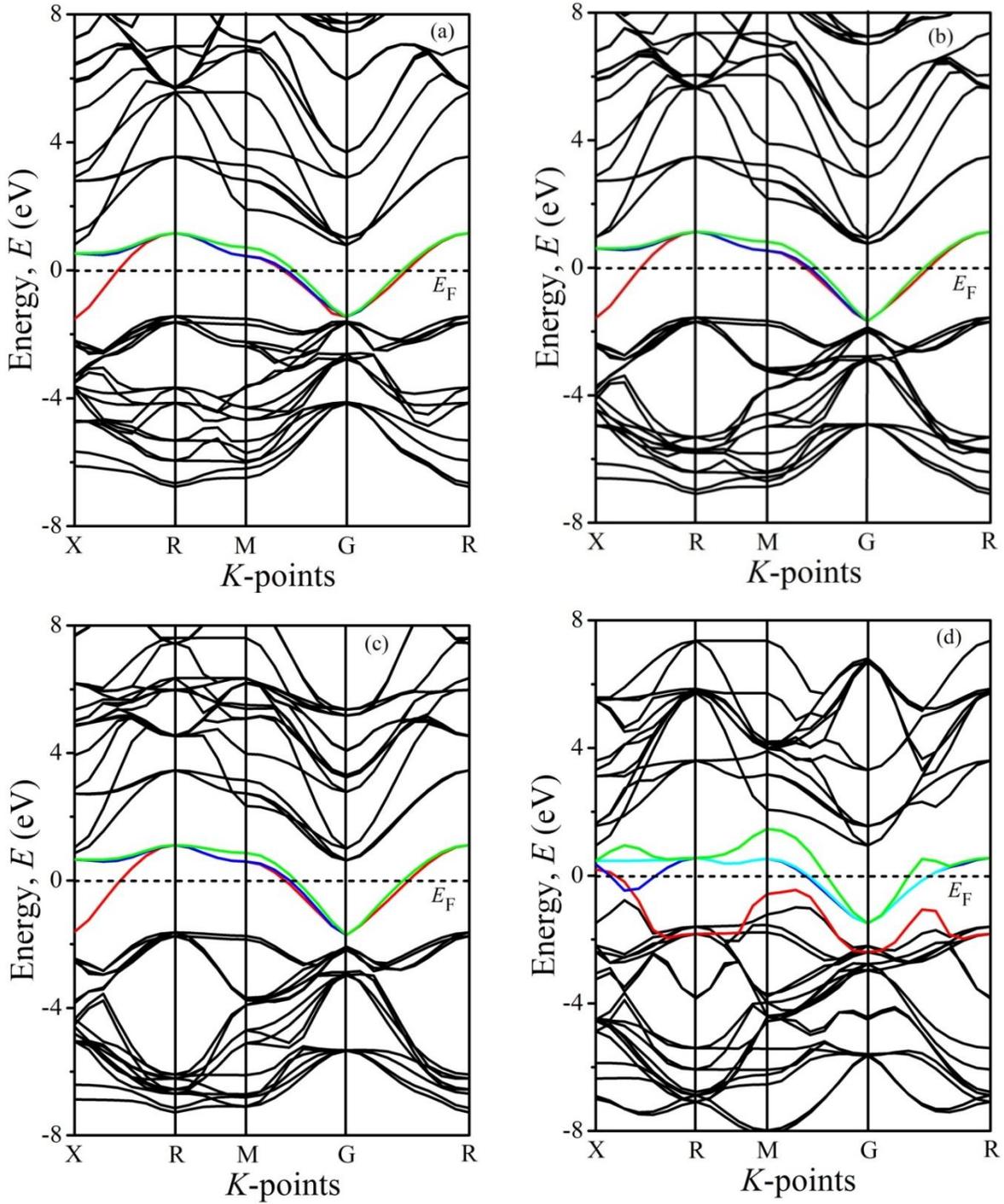

Fig. 3. (Color online) Calculated band structures of (a) BaVO$_3$ (b) SrVO$_3$ (c) CaVO$_3$ and (d) PbVO$_3$ along the high symmetry directions in the Brillouin zone at ambient conditions.

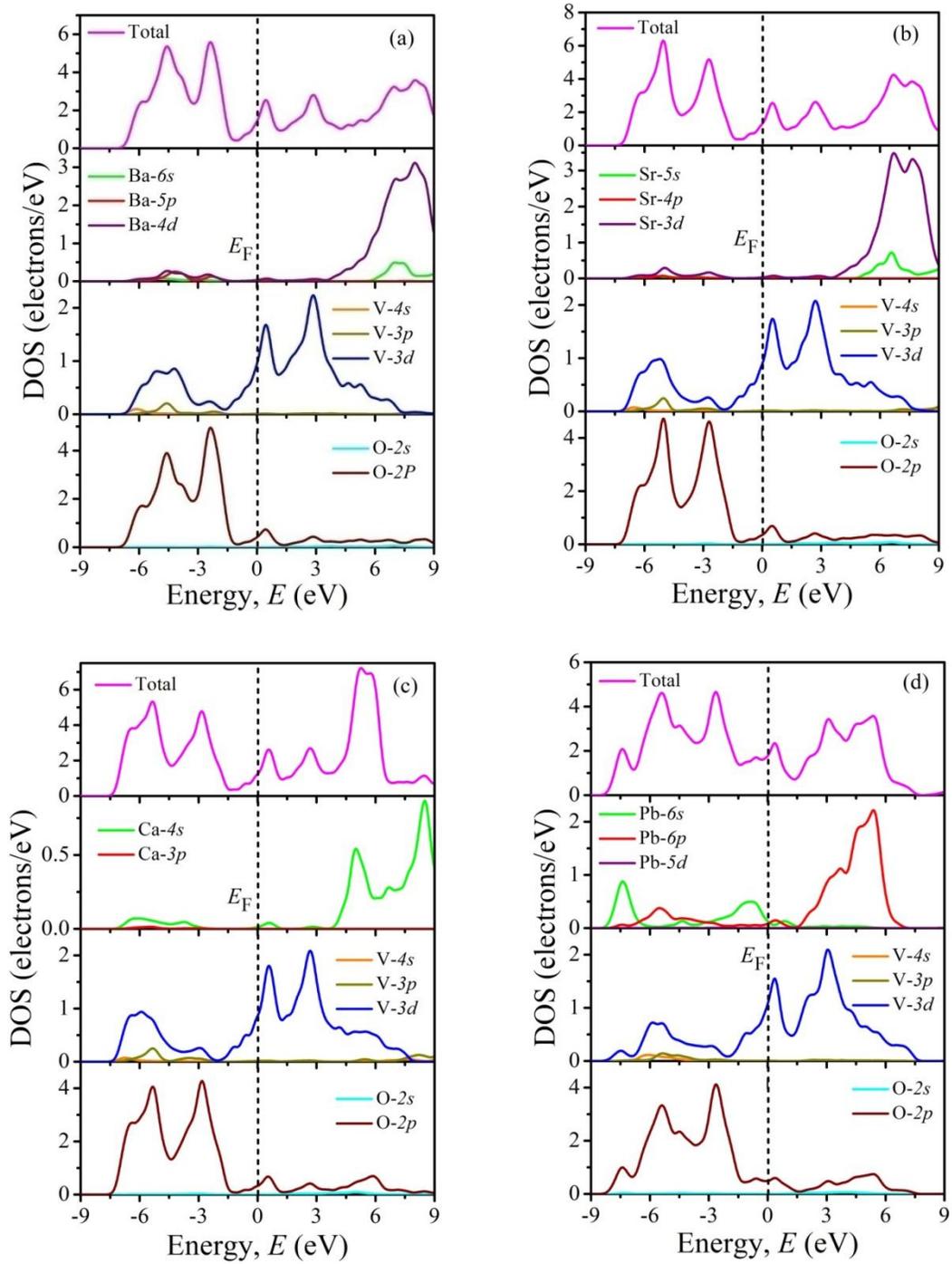

Fig. 4. (Color online) Total and partial electron energy density of states of (a) BaVO$_3$ and (b) SrVO$_3$ (c) CaVO$_3$ and (d) PbVO$_3$.

*3.3.2. Electronic charge density and Fermi surface*

The valence electronic charge density maps (in the units of e/Å$^3$) have been depicted in Fig. 4 to understand the distribution of the total electronic charge density of AVO$_3$ materials. The scale in the right side represents the intensity of electron density. The blue color presents the light density of electron whereas the red color shows high density of electron. The O–V bond in AVO$_3$ coincides with the hybridization between V–3$d$ and O–2$p$ orbitals in DOS (Fig. 4). Since, the charge density distribution is essentially spherical around all the atoms for all the compounds under this study show ionic nature. The ionic nature is an effect of metallic characteristics [79] which indicates the O–V bonds in AVO$_3$ manifest metallic nature. Therefore, a strong isotropic combination of chemical bonds such as ionic and metallic interactions exists from this calculation for AVO$_3$ compounds.

The Fermi surfaces of AVO$_3$ materials for the bands (colored lines) crossing the Fermi level are shown in Fig. 6. The Fermi surface topologies of BaVO$_3$, SrVO$_3$ and CaVO$_3$ compounds are almost similar but partially different for

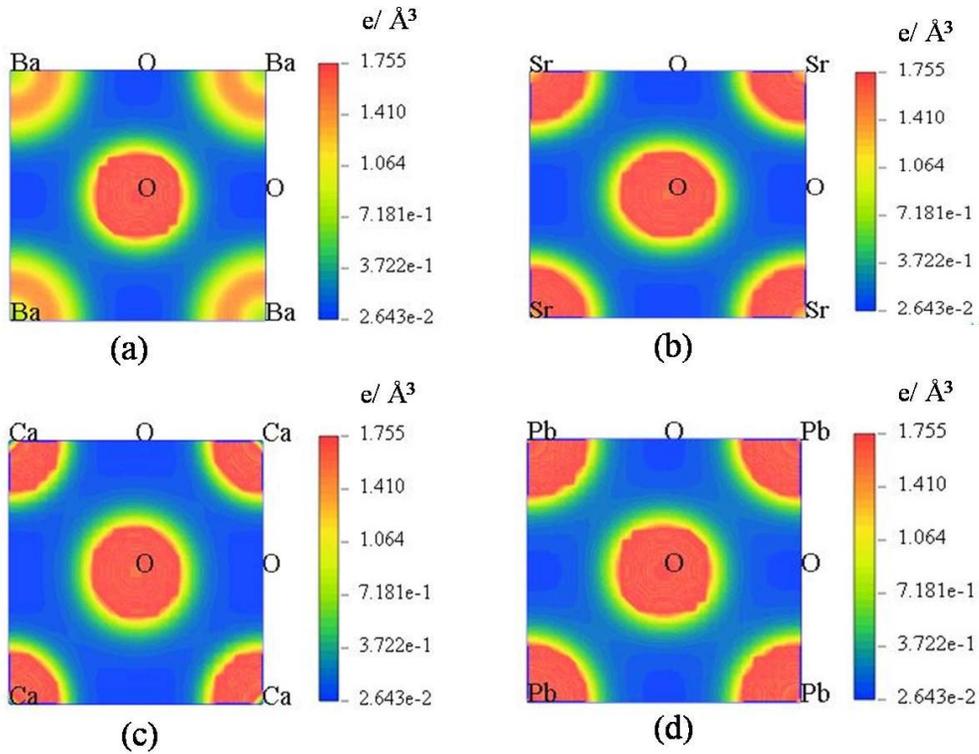

Fig. 5. (Color online) Electronic charge density of (a) BaVO$_3$ (b) SrVO$_3$ (c) CaVO$_3$ and (d) PbVO$_3$.

PbVO$_3$. From the topology we can see that there is a hole−like Fermi surface look like cylindrical cross sections with six windows surrounded at the G−point. A hole pocket also exists around X−point connected with the hole−like

Fermi surface surrounding the G–point. At the corner of the Brillouin zone, there is one electron–like Fermi surface around R–point not connected with the hole like Fermi surface for BaVO$_3$, SrVO$_3$ and CaVO$_3$ but connected for PbVO$_3$. Therefore, it is evident that both electron and hole–like Fermi surfaces are present in all compounds under study which predict the multiple–band nature of AVO$_3$.

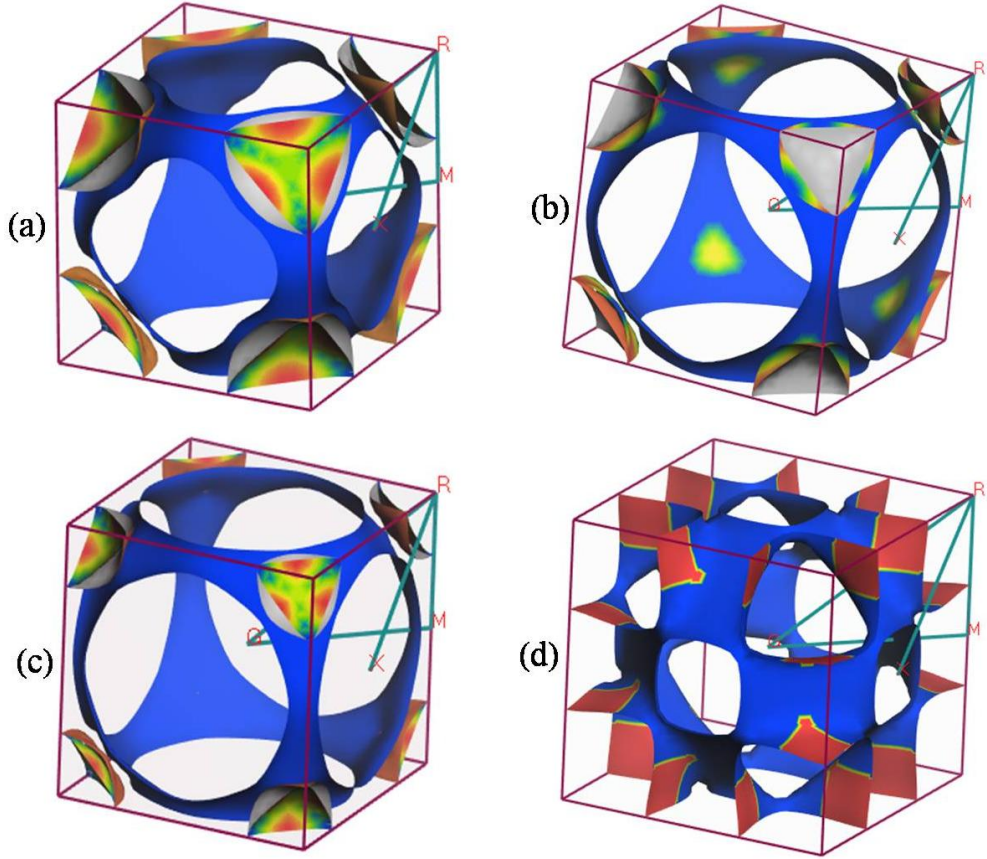

Fig. 6. (Color online) Fermi surface topology of (a) BaVO$_3$ (b) SrVO$_3$ (c) CaVO$_3$ and (d) PbVO$_3$.

*3.4. Optical Properties*

Optical properties of a material are closely related to the material response to incident electromagnetic radiation. The response to visible light is particularly important from the view of optoelectronic applications. The response to incident radiation is completely determined by various energy dependent parameters, namely dielectric function, refractive index, loss function, conductivity, reflectivity and absorption coefficient.

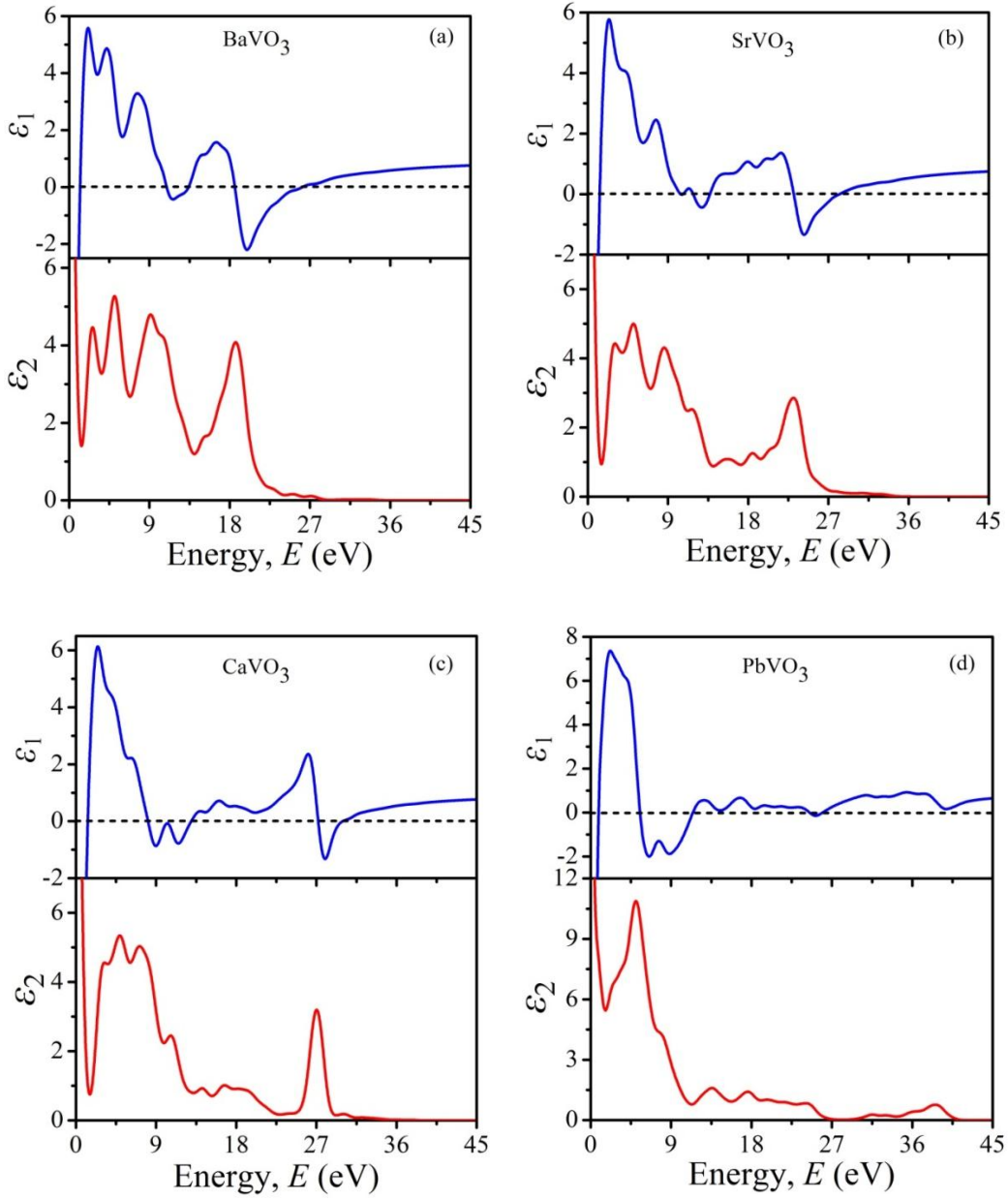

Fig. 7. (Color online) Energy dependent dielectric function (real part, $\epsilon_1$ and imaginary part, $\epsilon_2$) of (a) BaVO$_3$ (b) SrVO$_3$ (c) CaVO$_3$ and (d) PbVO$_3$.

The real ($\varepsilon_1$) and imaginary ($\varepsilon_2$) parts of the dielectric function of AVO$_3$ materials are shown in Fig. 7(a–d). Since AVO$_3$ compounds are metallic in nature, we employ the Drude plasma frequency of 3 eV and damping factor of 0.05 eV to investigate dielectric properties. In this investigation, the peak position of the real part of the dielectric function is connected to the electron excitation and peak is mainly arisen due to the intraband transitions. From Fig.

6 it is apparent that $\varepsilon_1(\omega)$ shows a peak at around 1 eV, which is related to the intraband transitions. For metal and metal-like systems, it is well known that in the intraband contributions from the conduction electrons mainly in the low-energy infrared are the part of the spectra. The $AVO_3$ compounds here are metallic in nature on the basis of the electronic band structure. So, the intraband contributions in $AVO_3$ come from the conduction electrons. It is significant that the $\varepsilon_2(\omega)$ reaches zero at around 30 eV in the ultraviolet energy region, which indicates the materials $AVO_3$ are transparent and optically anisotropic as well. The anisotropic nature of $AVO_3$ materials were also observed by the elastic properties.

The refractive index ($n$) denotes the phase velocity, when the electromagnetic wave (as light) passes through the material, displayed in Fig. 8(a). The calculated values of the static refractive index $n(0)$ is found to be 5.61, 6.25, 5.92 and 5.21 respectively for $BaVO_3$, $SrVO_3$, $CaVO_3$ and $PbVO_3$ and decreases in the high energy region.

The information about the solar energy conversion efficiency is provided by absorption coefficient ($\alpha$) indicates the penetration of a specific light energy into the material up to absorption [80]. From Fig. 8(b), the absorption spectra starts with zero value, also increases with increase in energy and finally decreases with further increase in energy for all phases. In general, the intraband contribution to the optical properties produces the low energy infrared part of the spectra. On the other hand, the peaks in the high energy region of the absorption and conductivity spectra may arise due to the interband transition. Notably, it is visual from Fig. 8(b) and 8(c) that the variation of the conductivity spectra is similar to the absorption spectra. So, we may conclude that the photoconductivity of $AVO_3$ increases as a result of absorbing photons [81].

The real part of photoconductivity ($\sigma$) spectra of $AVO_3$ materials as shown in Fig. 8(c) starts with zero photon energy is a sign of zero band gap for all the materials which is clearly seen in the electronic band structure calculations (Fig. 3) also. The values of maximum photoconductivity are 9.22, 7.98, 10.44 and 6.81 found at 18.75, 23.15, 27.06 and 5.31 eV for $BaVO_3$, $SrVO_3$, $CaVO_3$ and $PbVO_3$ respectively.

The reflectivity ($R$) spectrums of $AVO_3$ are displayed in Fig. 8(d). A higher reflectivity is seen in the infrared region approximately 49, 53, 51 and 46 % of the total radiation for $BaVO_3$, $SrVO_3$, $CaVO_3$ and $PbVO_3$ respectively and in the high energy region some peaks arisen due to inter band transition. Note here that, maximum reflectivity spectra were achieved in case of $SrVO_3$ compound. However, the high reflectivity spectra of these compounds indicate that these materials could be used as a promising coating material to diminish solar heating. The high value of reflectivity in low energy region reveals the characteristics of high conductance in the low energy region [82].

The photon energy loss spectra ($L$) of $AVO_3$ are shown in Fig. 8(e). The energy loss function is a significant index to express the loss of energy of a fast electron when it passes through a material [83]. In the graph of loss function, the peaks are connected with the plasma resonance and the frequency associated with it is defined as the plasma frequency $\omega_p$ [84]. The highest peaks are found at about 26.15, 28.34, 29.79 and 26.07 eV for $BaVO_3$, $SrVO_3$, $CaVO_3$ and $PbVO_3$ respectively, which reveal the plasma frequency of them and these values are very analogous to the previously reported article [41]. If the frequency of incident light is higher than the plasma frequency, the materials $AVO_3$ become transparent.

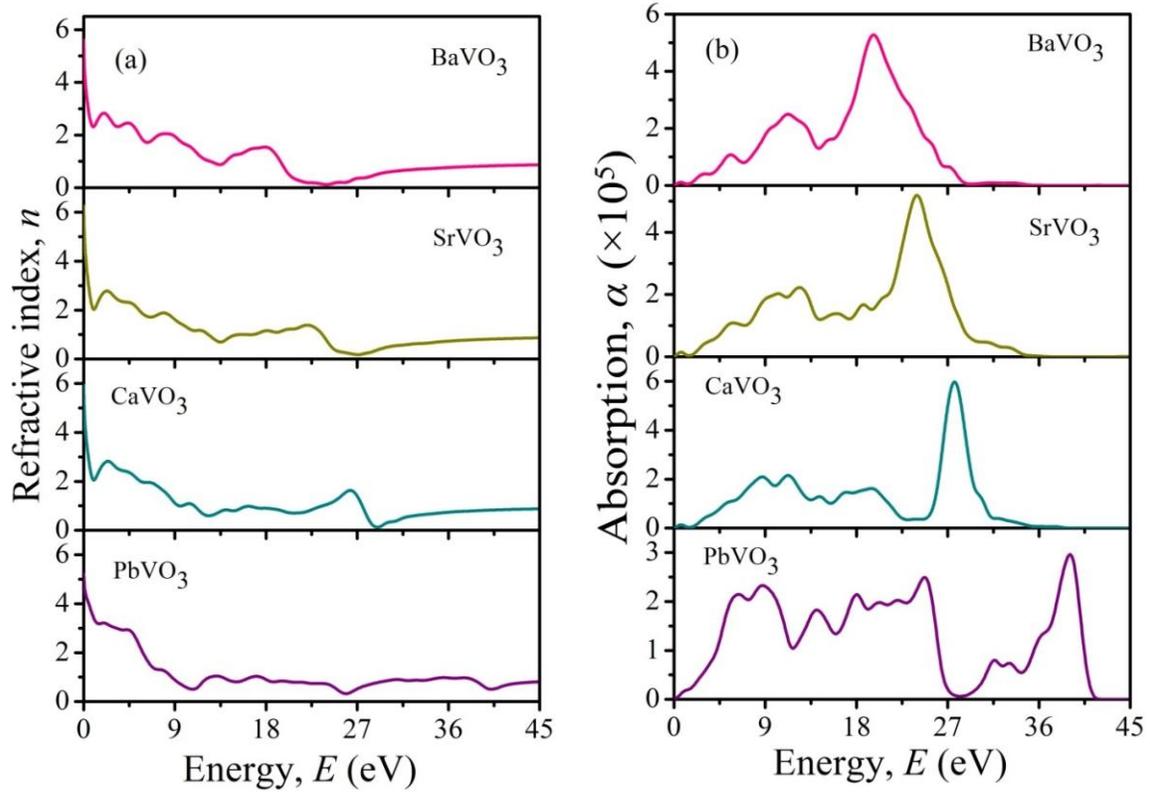

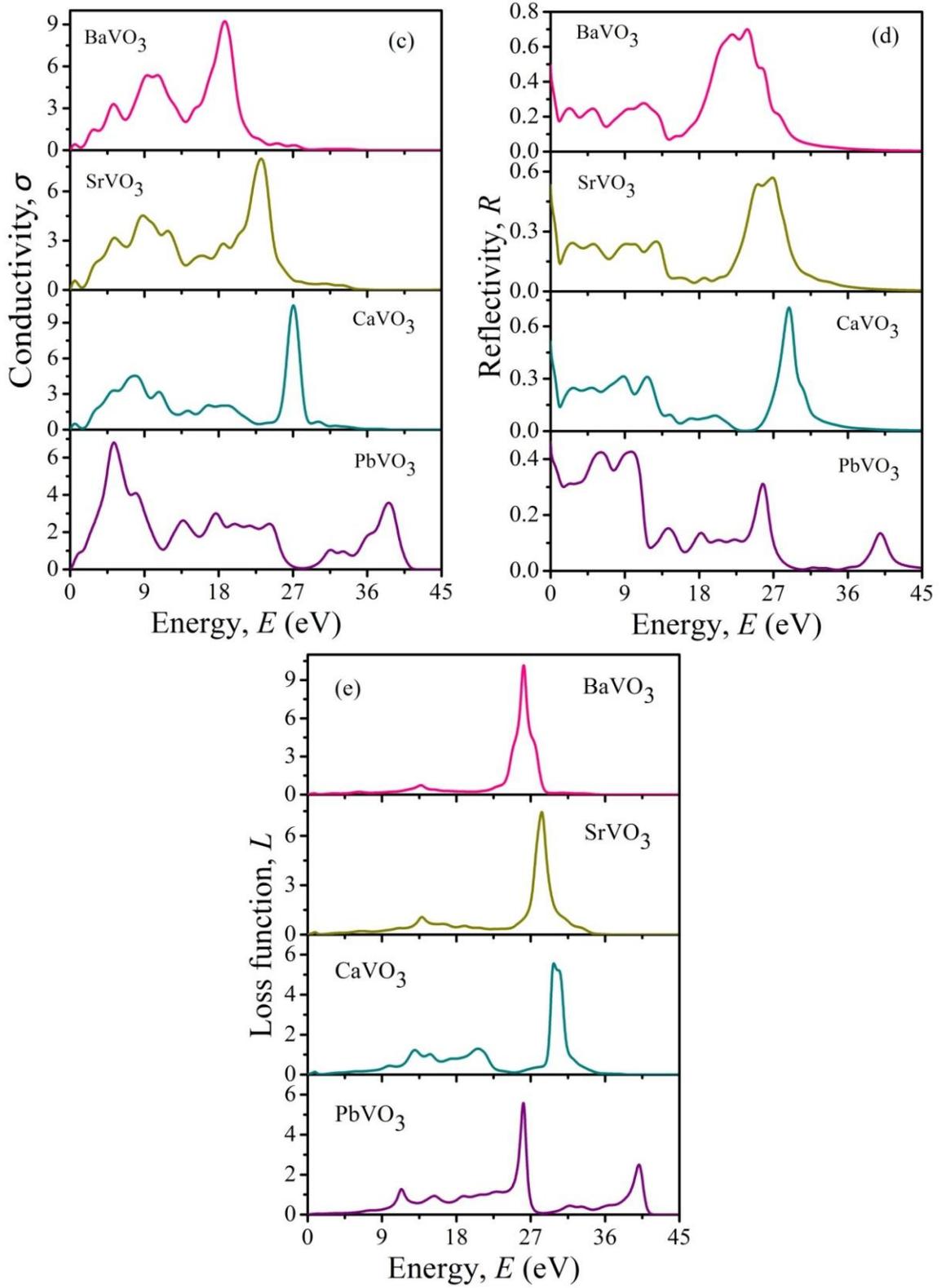

Fig. 8. (Color online) Energy dependent of (a) Real part of refractive index, $n$, (b) absorption, $\alpha$, (c) real part of conductivity, $\sigma$, (d) reflectivity, $R$ and (e) loss function, $L$ of the $AVO_3$ materials.

*3.5. Population analysis*

To understand the chemical bonding nature in solids, Mulliken atomic population analysis provides many interesting information [85]. Mulliken orbitals and Mulliken populations obtained as well as the estimated overlap populations for nearest neighbour atoms of $AVO_3$ are listed in Table 3 and Table 4 respectively. We can see from Table 3 that A (= Ba, Sr, Pb, Ca) and V atoms carry only positive charges while O atoms carry negative charges indicating that charge sharing occurs from A and V to O atoms. In the bond overlap population, zero value indicates a perfectly ionic bond and the values higher than zero indicate the increasing levels of covalency [86]. From Table 4, the atomic bond populations for all phases are positive and greater than zero which reveals the covalent nature of these compounds. Thus the higher values of the overlap population of all bonds imply covalent bonding of $AVO_3$. It is noted from Table 4 that O−V bond in $PbVO_3$ is more covalent than the other phases. Therefore, the strong covalent bond creates a significant resistance which acts against the initialization of plastic deformation and pin the dislocation, resulting in a quite large hardness.

*3.6. Phonon properties*

The phonon dispersion curve (PDC) of a material gives some interesting information regarding structural stability and vibrational contribution in the thermodynamic properties such as thermal expansion, Helmholtz free energy, and heat capacity [87,88]. First principles finite displacement phonon dispersion spectra for $AVO_3$ compounds have been calculated using the norm−conserving pseudo potential method. The calculated PDC along the principal symmetry direction of the Brillouin zone (BZ) are presented in Fig. 9. It is established that the total number of phonon modes are strictly given by three times the number of atoms existed in a primitive cell. Since the primitive cell of $AVO_3$ consists of 5 atoms, we found 15 phonon branches that comprise 3 acoustic modes and 12 optical modes. The PDCs of $AVO_3$ decompose into two separate parts because of their significant mass differences of atoms. The A and V atoms can be predominated in low frequency region and the vibrations of the light O atoms in high frequency region. A compound is identified as dynamically stable if the phonon frequencies for all wave vectors are positive. The instability of the compound having imaginary frequency i.e., phonon dispersion happens below the zero−frequency line.

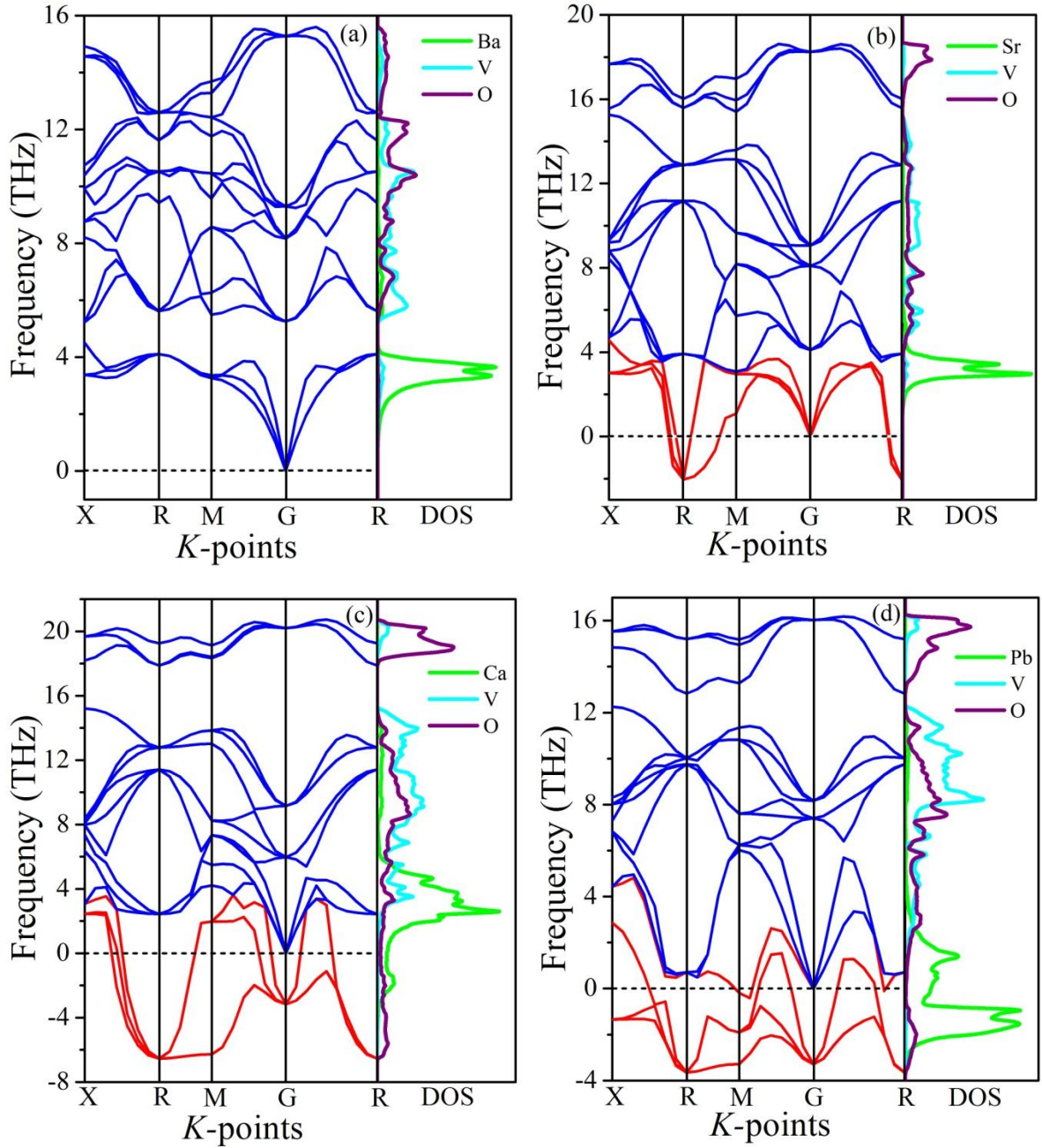

Fig. 9. Calculated phonon spectra (phonon dispersion in the left and phonon partial DOS in the right) of (a) $BaVO_3$ (b) $SrVO_3$ (c) $CaVO_3$ and (d) $PbVO_3$. The dashed line is at zero phonon frequency.

As we see from the Fig. 9(a), $BaVO_3$ shows stability but the compounds $SrVO_3$, $CaVO_3$ and $PbVO_3$ [Fig. 9(b–d)] have imaginary phonon branches (shown by red lines), hence showing dynamical instability. For $SrVO_3$, the unstable phonon is observed at R point, while the unstable phonons are founds at R, M, and G points in $CaVO_3$ and PbVO$_3$ compounds. In CaVO$_3$ and PbVO$_3$, the most unstable phonon found at R point and the energy of phonon depends slightly on the wave vector along the R–M line. The results are almost similar to those observed in CaTiO$_3$ and PbTiO$_3$ ferroelectrics [41]. The unstable phonons with wave vectors near the edges of the cubic Brillouin zone correspond to oxygen octahedral rotation and explain structural distortions [1]. In SrTiO$_3$, it is reported that the unstable transverse optic (TO) phonon at the G point corresponds to ferroelectric distortions related to phase transitions, where Ti atoms move parallel to one of the Ti–O bonds and the oxygen octahedra moves in the opposite direction [1,89]. Albeit the most unstable phonon is found at R point in CaVO$_3$ and PbVO$_3$, TO mode is also observed at the G point and the R and the M are negligible dispersion points. Therefore, the modes correspond to imaginary frequencies of AVO$_3$ (A = Sr, Ca, Pb) might be related to the structural transitions. Recently, the lower symmetry tetragonal phase with *P4mm* symmetry of PbVO$_3$ was also reported [12–14]. In order to have a better understanding the instability of the AVO$_3$, we also studied the tolerance factor (t) of these compounds. The tolerance factor of a perovskite compound can be estimated by Goldschmidt formula [90] : t = $(r_A + r_0)/\sqrt{2}(r_B + r_0)$, where $r_A$ is the ionic radius of the A–cation, $r_B$ is the ionic radius of the B–cation and $r_0$ is the ionic radius of the anion. The ionic radius of cations and anions is reported [91] elsewhere. Using this equation, the calculated values of tolerance factor of BaVO$_3$, SrVO$_3$, CaVO$_3$ and PbVO$_3$ are 0.884, 0.833, 0.775 and 0.836, respectively. The value of tolerance factor for ideal cubic structure is 1.00. The value of t becomes smaller than 1.00, if the A–cation is smaller than the ideal value. Hence, the octahedra will tilt in order to fill space. However, the cubic structure belongs to 0.89 < t <1.00 [92,93]. The lower value of t will lower the symmetry of the crystal structure. It is worth noting that, the value of t for BaVO$_3$ is very close to the lower limit of the range of cubic perovskite structure; give rise to dynamical stability of the BaVO$_3$. On the other hand, the value of t is deviated from the boundary condition of the cubic perovskite, results in an instability of the AVO$_3$ (A= Sr, Ca, Pb). Therefore, it is suggested that the instability of the modes of AVO$_3$ (A = Sr, Ca, Pb) can be due to octahedra tilting/rotation and related to the phase transitions of these compounds.

For better understanding, atomic contribution of the PDC of AVO$_3$ compound, we study the phonon partial DOS (PHDOS). The PHDOS is depicted in Fig. 9 side by side with the PDC of the bands by comparing corresponding peaks. The peaks in PHDOS arise due to the relative flatness of the bands. On the contrary, both the downward and upward sloping bands reduce the peak–heights in the total DOS. It can be well clarified by considering a peak and the flatness/sloping of its corresponding bands. It is clearly seen from Fig. 9, in the low–frequency regions (about <

4 THz), the PDC are mainly dominated by the vibration of A (=Ba, Sr, Ca, Pb) atom, while V atom has little contribution in CaVO$_3$ and PbVO$_3$ compounds. The frequency region of 4~13 THz, the PDC comes from the vibrations of V and O atoms, whereas the higher frequency regions (>13 THz) is mostly dominated by the vibrations of O atoms.

*3.7. Thermodynamic Properties*

The thermodynamically potential functions namely Helmholtz free energy $F$, energy $E$, entropy $S$, specific heat $C_v$, and Debye Temperature $\Theta_D$ of AVO$_3$ are calculated at zero pressure using the quasi–harmonic approximation [94]. The expressions used to calculate $F$, $E$, $S$, $C_v$ and $\Theta_D$ can be found elsewhere [95] based on the work by Baroni et al. [96]. Fig. 10 (a–c) represents the calculated $F$, $E$, $S$, $C_v$ and $\Theta_D$ of the AVO$_3$ compounds within the temperature range from 0 to 1000 K in which the harmonic model is assumed to be valid. We see from the Fig. 10 (a) that the enthalpy, entropy and free energy have almost zero values at a temperature lower than 100 K. But it is evident from Fig. 10 (a) that the temperature dependence of Helmholtz free energy ($F$) is found to decrease gradually with increase in temperature. This decreasing trend of $F$ is very common and it becomes more negative during any natural process. The difference between internal energy of a system and the amount of energy that can not be used to perform work is defined as free energy. This unusable energy is expressed as the product of entropy of a system and the absolute temperature of the system. We present the entropy as $TS$ in order to compare Helmholtz free energy and internal energy. Here, $S$ is the lattice entropy resulting from lattice vibration which rises with increasing temperature. We can also see from the Fig. 10(a) that, the variation of internal energy ($E$) with temperature reveals an increasing trend as expected for solids. Fig. 10(b) displays the curves of specific heat $C_v$ of AVO$_3$ compounds. As temperature increases phonon thermal softening occurs and as a result the heat capacities increase with increasing temperature. As we can see from the Fig. 10(b) that, in the low temperature limit, $C_v$ of AVO$_3$ follow the Debye model which is proportional to $T^3$, as expected [97]. These curves follow Dulong–Petit law at high temperatures [98]. Above 400 K, $C_v$ increases slowly with temperature and gradually approaches the Dulong–Petit limit for all phases of AVO$_3$. Fig. 10(c) also shows the temperature dependence of Debye temperature $\Theta_D$ which is calculated from PHDOS. It is observed that $\Theta_D$ increases with increasing temperature which indicates the change of the vibration frequency of particles under temperature effects. This is also related to the bonding strength. The weaker the bonds in solids, the lower the value of $\Theta_D$, and the heat capacity reach the classical Dulong–Petit value at a lower temperature.

The other interesting thermodynamic properties such as Debye temperature ($\Theta_D$), melting temperature ($T_m$) and minimum thermal conductivity ($K_{min}$) of the AVO$_3$ materials have been investigated for the first time to understand the behaviour of this material under high temperatures as well as high pressures. The $\Theta_D$ is an essential parameter of solids to discuss some interesting physical process related to lattice vibrations, melting point, specific heat, thermal conductivity etc. [99].

Table 3. Mulliken atomic population analysis of AVO$_3$ compounds.

| Compounds | Species | Mulliken atomic populations | | | | Charge (e) |
|---|---|---|---|---|---|---|
| | | s | p | d | Total | |
| BaVO$_3$ | Ba | 2.13 | 5.97 | 0.79 | 8.88 | 1.12 |
| | V | 0.20 | 0.47 | 3.23 | 3.90 | 1.10 |
| | O | 1.85 | 4.89 | 0.00 | 6.74 | -0.74 |
| SrVO$_3$ | Sr | 2.15 | 5.99 | 0.71 | 8.85 | 1.15 |
| | V | 0.19 | 0.47 | 3.25 | 3.90 | 1.10 |
| | O | 1.85 | 4.90 | 0.00 | 6.75 | -0.75 |
| CaVO$_3$ | Ca | 2.15 | 5.99 | 0.54 | 8.69 | 1.31 |
| | V | 0.18 | 0.54 | 3.27 | 4.00 | 1.00 |
| | O | 1.85 | 4.92 | 0.00 | 6.77 | -0.77 |
| PbVO$_3$ | Pb | 3.59 | 7.09 | 10.00 | 20.69 | 1.31 |
| | V | 0.39 | 0.44 | 3.24 | 4.07 | 0.93 |
| | O | 1.86 | 4.89 | 0.00 | 6.75 | -0.75 |

Table 4. Calculated Mulliken bond number $n^\mu$, bond length $d^\mu$ and bond overlap population $P^\mu$, of AVO$_3$ compounds.

| Compounds | Bonds | $n^\mu$ | $d^\mu$(Å) | $P^\mu$ |
|---|---|---|---|---|
| BaVO$_3$ | O–V | 3 | 1.95453 | 0.88 |
| SrVO$_3$ | O–V | 3 | 1.91349 | 0.86 |
| CaVO$_3$ | O–V | 3 | 1.89015 | 0.89 |
| PbVO$_3$ | O–V | 3 | 1.94337 | 1.01 |

Using the average sound velocity, the $\Theta_D$ can be calculated by the following equation [99],

$$\theta_D = \frac{h}{k_B}\left[\frac{3m}{4\pi}\left(\frac{N_A\rho}{M}\right)\right]^{\frac{1}{3}} v_m$$

Where, $h$ and $k_B$ are the Planck's and Boltzmann constants, respectively. $V$ is the volume of unit cell, $n$ is the number of atoms within a unit cell, and $v_m$ is the average sound velocity. The $v_m$ in the crystal is calculated by the following equation,

$$v_m = \left[\frac{1}{3}\left(\frac{2}{v_t^3} + \frac{1}{v_l^3}\right)\right]^{-\frac{1}{3}}$$

Here, $v_l$ and $v_t$ are the longitudinal and transverse sound velocities, respectively. With the help of bulk modulus, $B$ and shear modulus, $G$, the $v_l$ and $v_t$ can be determined using the following equations,

$$v_l = \left(\frac{B+\frac{4}{3}G}{\rho}\right) \text{ and } v_t = [G/\rho]^{1/2}$$

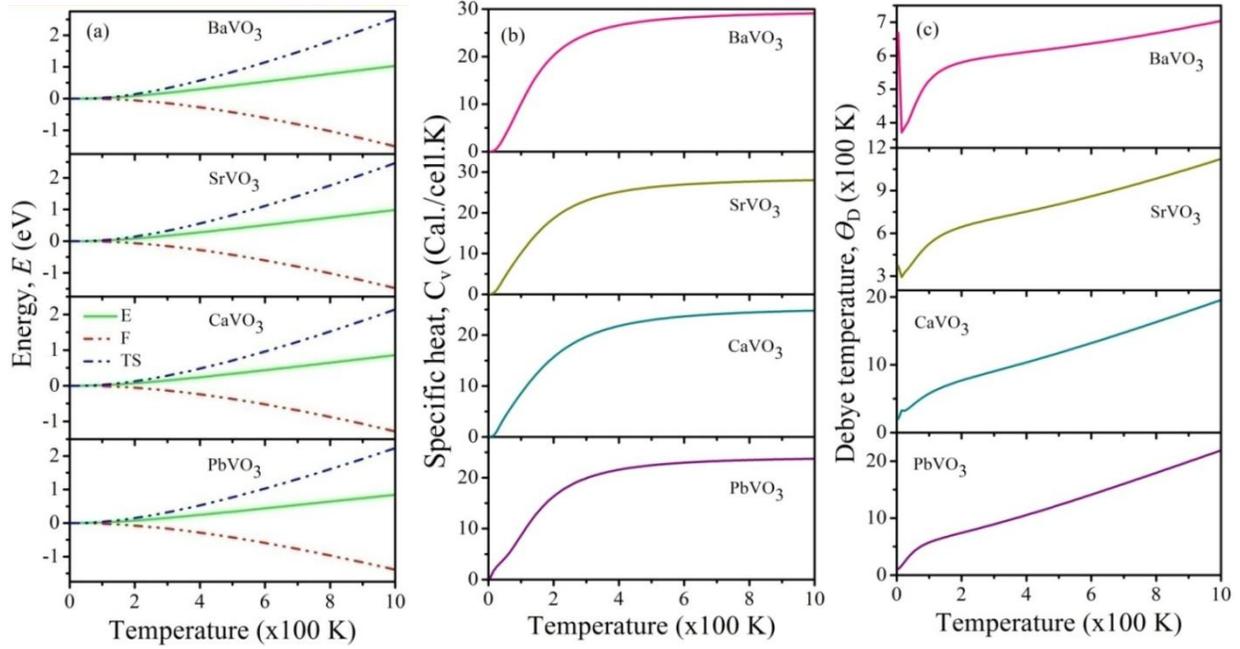

Fig. 10. Temperature dependence of the thermo dynamical potential functions of $AVO_3$.

We also calculated melting temperature ($T_m$) of the $AVO_3$ compound via the following empirical formula using elastic constants $C_{ij}$

[100]: $T_m = 354 + \frac{4.5(2C_{11}+C_{33})}{3}$

Since, the axial lengths are equal in cubic structure, thus $C_{11}$, $C_{22}$ and $C_{33}$ are equal. Moreover, we calculated another important parameter such as thermal conductivity used to investigate the heat conduction of a material. It is well established that the minimum thermal conductivity is directly concerned to the temperature while the temperature of a material gradually increased the conductivity of the material then gradually decreased to a certain value [101]. Although many similar equations are available to calculate the minimum thermal conductivity, but in this report the entity $K_{min}$ has been calculated by using the Clarke expression [102] and can be denoted as,

$$K_{min} = K_B \, v_m \, (M/n\rho \, N_A)^{-2/3}$$

Where, $K_B$ is the Boltzmann constant, $v_m$ is the average sound velocity, $M$ is the molecular mass, $n$ is the number of atoms per molecule and $N_A$ is the Avogadro's number used for the calculation.

The calculated values of $\Theta_D$, $T_m$, $K_{min}$, $v_m$, $v_t$ and $v_l$ for all the compounds under this study using above equations are listed in Table 5. In general, a higher Debye temperature involves in a higher phonon thermal conductivity and vice–versa. The relatively high value of $\Theta_D$ and $K_{min}$ of BaVO$_3$, SrVO$_3$ and CaVO$_3$ imply the high thermal conductivity and they might not be used as thermal barrier coating (TBC) materials but PbVO$_3$ has lower $\Theta_D$ and $K_{min}$ and convenient to exploit as thermal barrier coating (TBC) materials [32,65].

Table 5. The Calculated density ($\rho$), longitudinal, transverse, and average sound velocities ($v_l$, $v_t$, and $v_m$) Debye temperature ($\Theta_D$), melting temperature ($T_m$) and minimum thermal conductivity ($K_{min}$) of AVO$_3$ compounds.

| Compounds | $\rho$ (g/cm$^3$) | $v_l$ (km/s) | $v_t$ (km/s) | $v_m$ (km/s) | $\Theta_D$ (K) | $T_m$ (K) | $K_{min}$ (Wm$^{-1}$K$^{-1}$) |
|---|---|---|---|---|---|---|---|
| BaVO$_3$ | 6.566 | 7.188 | 4.185 | 4.642 | 608 | 1807 | 1.22 |
| SrVO$_3$ | 5.526 | 7.916 | 4.501 | 5.003 | 665 | 1956 | 1.37 |
| CaVO$_3$ | 4.272 | 8.868 | 4.886 | 5.593 | 753 | 2005 | 1.57 |
| PbVO$_3$ | 8.656 | 5.811 | 3.077 | 3.366 | 441 | 1654 | 0.63 |

## 4. Conclusions

In the present study, we have performed first principle calculations to investigate the physical properties of AVO$_3$ materials using the density functional theory dependent CASTEP code software package. First time we have calculated and discussed the elastic constants, Bulk modulus, shear modulus, Young's modulus, Vicker hardness, peierls stress, elastic anisotropy, band data, DOS, optical properties, Mulliken population analysis, phonon spectra and thermodynamic properties of cubic phase AVO$_3$. The calculated lattice parameters are in good agreement with the available experimental values demonstrating the reliability of these calculations. The calculated elastic constants also satisfy the mechanical stability conditions for all compounds under this study. The Poisson's and Pugh's ratio reveal the brittle nature of BaVO$_3$ and SrVO$_3$ whereas ductile behaviour of CaVO$_3$ and PbVO$_3$. Interestingly, all the compounds show high Vicker hardness. Moreover, the Universal anisotropy index shows the negligible elastic anisotropy of the cubic AVO$_3$. All the compounds reveal intermediate values of Peierls stress and is confirmed that the dislocation movements in AVO$_3$ do not occur as easily as some selected MAX phases and inverse perovskites. The AVO$_3$ compounds show metallic character and main contribution comes from the O−2p and V−3d orbitals at Fermi level in DOS for all compounds. Furthermore, the multiple−band nature of AVO$_3$ compounds is evident from the existence of both electron and hole−like Fermi surfaces. The bonding properties reveal that all the compounds possess intra atomic bonding with a mixture of ionic, covalent and metallic interactions. The optical properties of these compounds are investigated that exhibit interesting results. The study of optical property (−reflectivity spectra) of these compounds indicates that all the materials could be used as a promising coating material to reduce solar heating. The calculated phonon dispersion curves confirm the dynamical stability only for BaVO$_3$ among the four materials under this investigation. Though a decrease is observed for the free energy, an increase occurs for the enthalpy and entropy with increasing temperature above 100 K. Debye temperature, melting temperature and minimum thermal conductivity also calculated and all the parameters show reasonably high values. From the calculated values of Debye temperature and minimum thermal conductivity it is predicted that among the four materials under this work only PbVO$_3$ has potential to be used as thermal barrier coating (TBC) material. Finally, we hope that, the present study will motivate to investigate the different properties of others laves phase materials.

## 5. Acknowledgements

This research work was jointly supported by grant number 12210280 from the faculty of engineering, university of Rajshahi and ministry of national science and technology (NST), Bangladesh.